\documentclass[twocolumn,showpacs,preprintnumbers,amsmath,amssymb]{revtex4}

\usepackage{amsmath,amssymb,graphicx}
\usepackage{graphicx}
\usepackage{dcolumn}
\usepackage{bm}
\usepackage{epsfig}
\usepackage{here}
\usepackage{float}
\usepackage{amsmath}
\usepackage[usenames]{color}

\newcommand{\bea}{\begin{eqnarray}}
\newcommand{\eea}{\end{eqnarray}}

\begin{document}
\draft


\title{Constraints on a $f(R)$ gravity dark energy model with early
       scaling evolution}
\author{Chan-Gyung Park${}^{1}$, Jai-chan Hwang${}^{1}$,
        and Hyerim Noh${}^{2}$}
\address{${}^{1}$Department of Astronomy and Atmospheric Sciences,
                 Kyungpook National University, Taegu, Korea \\
         ${}^{2}$Korea Astronomy and Space Science Institute,
                 Daejon, Korea}


\begin{abstract}

The modified gravity with $f(R)=R^{1+\epsilon}$ ($\epsilon>0$)
allows a scaling solution where the density of gravity sector
follows the density of the dominant fluid. We present initial
conditions of background and perturbation variables during the
scaling evolution regime in the modified gravity. As a possible dark
energy model we consider a gravity with a form
$f(R)=R^{1+\epsilon}+qR^{-n}$ ($-1<n \le 0$) where the second term
drives the late-time acceleration. We show that our $f(R)$ gravity
parameters are very sensitive to the baryon perturbation growth and
baryon density power spectrum, and present observational constraints
on the model parameters. Our analysis suggests that only the
parameter space extremely close to the $\Lambda\textrm{CDM}$ model
is allowed.

\end{abstract}

\noindent \pacs{98.80.-k, 95.36.+x}

\maketitle

\section{Introduction}
\label{sec:intro}

The current accelerated expansion of the universe has been theoretically
explained by either the cosmological constant or some kind of dynamical
dark energy in the context of field or modified gravity (see Ref.\
\cite{DE-review-2010} for recent reviews).

In the modified gravity such as $f(R)$ gravity \cite{fR-gravity-DE}
(see Refs.\ \cite{fR-gravity-review} for reviews and
\cite{fR-gravity-recent,Amendola-etal-2007,Li-Barrow-2007,
Amendola-Tsujikawa-2008,Hu-Sawicki-2007,Starobinsky-2007,Appleby-Battye-2007,
Tsujikawa-2008,Amendola-etal-2007-PRL,Carloni-etal-2008,
Clifton-2008} for recent works), background evolution sometimes
confronts with a numerical difficulty. The background evolution is
not easily handled numerically during the early radiation dominated
era in the modified gravity with a functional form
$f(R)=R+f_\textrm{DE}(R)$, where $f_\textrm{DE}$ drives the
late-time acceleration. The cosmologically viable forms of
$f_\textrm{DE}$ proposed so far are $qR^{-n}$
\cite{Amendola-etal-2007,Li-Barrow-2007,Amendola-Tsujikawa-2008} and
forms given by Hu \& Sawicki \cite{Hu-Sawicki-2007}, Starobinsky
\cite{Starobinsky-2007}, and so on
\cite{Appleby-Battye-2007,Tsujikawa-2008}. In all models, the second
term $f_\textrm{DE}$ becomes extremely subdominant compared with $R$
in the early radiation dominated era so that the $f(R)$ gravity
effectively goes over into the Einstein gravity; in general,
however, the evolution could be more complicated, see
\cite{Clifton-2008}. Since the quantity $F \equiv df/dR$ becomes
extremely close to unity in such a situation, evolving a
differential equation like Eq.\ (\ref{eq:ddF}) or (\ref{eq:ddR})
below is sometimes not numerically feasible.

Adopting the modified form $f(R)=R^{1+\epsilon}+f_\textrm{DE}$ with
small positive $\epsilon$ together with appropriate initial
conditions we can evade this numerical problem (here we use the
Planck unit with $8\pi G\equiv 1 \equiv c$). It is known that the
first term $R^{1+\epsilon}$ which is dominant in the early epoch
allows the density of gravity sector to follow that of dominant
fluid (scaling evolution) \cite{Amendola-etal-2007}.

We are motivated to study the case in order to investigate the
observationally allowed regions with qualitatively different
evolution available in our case of $f(R)$ gravity. By considering
$R^{1+\epsilon}$ term, however, the gravity with
$f_\textrm{DE}=-2\Lambda$ does not go over into the Einstein gravity
in recent era. We will still consider values of $\epsilon$ which is
likely to be excluded by the solar-system test because with
vanishingly small $\epsilon$ the system of equations cannot be
handled numerically due to limited numerical precision in the early
era. Though, we will show that for smallest value of $\epsilon$ we
considered, the cosmological evolution we study is numerically
similar to the evolution in Einstein's gravity with cosmological
constant.

In this paper, we present initial conditions of background and perturbed
variables during the scaling regime in this gravity. Using the initial
conditions for scaling evolution we present background evolution, matter
(density) and cosmic microwave background (CMB) anisotropy power spectra,
and perturbation growth in the gravity with $f_\textrm{DE}=qR^{-n}$.
We show that the CMB power spectrum is not sensitive to the model parameters,
and explore the viable parameter space constrained by the type Ia supernova
(SNIa), matter power spectrum, and the future perturbation growth factor
observation.

Throughout this paper we assume spatial flatness ($K\equiv 0$).
Notations and the basic set of equations in $f(R)$ gravity are summarized
in Ref.\ \cite{Hwang-etal-2010}.

\section{Background and perturbation equations}

The background evolution in the $f(R)$ gravity is described by (see
Eqs.\ (43), (57) and (59) of Ref.\ \cite{Hwang-etal-2010})
\begin{equation}
   \ddot{F}+3H\dot{F}+\frac{1}{3}(2f-FR)=\frac{1}{3}(\mu_m-3p_m),
\label{eq:ddF}
\end{equation}
where $H\equiv \dot{a}/a$, $a(t)$ is the cosmic scale
factor, a dot represents a derivative with respect to the cosmic
time $t$, and $\mu_m$ and $p_m$ are collective density and pressure
for radiation ($R$) and matter ($M$): $\mu_m=\mu_R+\mu_M$ and
likewise for $p_m$. Using $\dot{F}=F_{,R}\dot{R}$ with $F_{,R}
\equiv dF/dR$, Eq.\ (\ref{eq:ddF}) is transformed into a
differential equation for $R$
\begin{equation}
   \ddot{R}+\frac{F_{,RR}}{F_{,R}}\dot{R}^2 + 3H\dot{R}+\frac{2f-RF}{3F_{,R}}
    =\frac{1}{3F_{,R}}(\mu_m-3p_m),
\label{eq:ddR}
\end{equation}
which is what we actually solved numerically. To evolve this
equation we need $H(t)$. For spatially flat Robertson-Walker metric,
we have (see Eqs.\ (43) and (57) of Ref.\ \cite{Hwang-etal-2010})
\begin{equation}
\begin{split}
   H^2 =\frac{1}{3F}\left(\mu_m + F\mu_X \right), \quad
   F\mu_X \equiv \frac{1}{2}(FR-f)-3H\dot{F},
\end{split}
\label{eq:H2}
\end{equation}
where $\mu_X$ indicates the energy density of $f(R)$ gravity sector
(hereafter $X$-component), and $R=6(2H^2+\dot H)$. With the
definition of conventional density parameters, $\Omega_i = \mu_i /
(3H^2)$ ($i=R$, $M$, $X$), we have a relation
$(\Omega_R+\Omega_M)/F+\Omega_X=1$ from Eq.\ (\ref{eq:H2}). The
equation of state of the $X$-component is defined as $w_X \equiv p_X
/ \mu_X$, where $p_X$ is the pressure of the $X$-component given by
(see Eq.\ (57) of Ref.\ \cite{Hwang-etal-2010})
\begin{equation}
   F p_X \equiv -\frac{1}{2}(FR-f)+\ddot{F}+2H\dot{F}.
\end{equation}

The perturbation equations in the CDM-comoving gauge (CCG), which
sets perturbed velocity of cold dark matter is set to zero ($v_c
\equiv 0$) as the temporal gauge (hypersurface) condition, are
presented in Eqs.\ (86)--(88), (66), and (67) of Ref.\
\cite{Hwang-etal-2010}. Using $\delta R$ as the perturbation
variable, we have
\begin{widetext}\bea
   && \mskip-36mu
       \left( {\delta R \over R} \right)^{\prime \prime}
       + \left( 3 + {\dot H \over H^2}
       + 2 {F_{,RR} \over F_{,R}} R^\prime
       + 2 {R^\prime \over R} \right)
       \left( {\delta R \over R} \right)^\prime
       + \Bigg\{
       {F \over 3 H^2 F_{,R}}
       - {R \over 3 H^2}
       + {F_{,RR} \over F_{,R}} \left[
       R^{\prime \prime} + \left( 3 + {\dot H \over H^2} \right)
       R^\prime \right] \nonumber \\
   &&  + {F_{,RRR} \over F_{,R}} R^{\prime 2}
       + {R^{\prime\prime} \over R}
       + {R^\prime \over R} \left( 3 + {\dot H \over H^2}
       + 2 {F_{,RR} \over F_{,R}} R^\prime \right)
       + {k^2 \over a^2 H^2} \Bigg\}
       \left( {\delta R \over R} \right)
       = {R^\prime \over R} \left( {\kappa \over H} \right)
       + {1 \over 3 H^2 F_{,R} R}
       \left( \delta \mu_m - 3 \delta p_m \right),
   \label{eq-fR-1} \\
   && \mskip-36mu
       \left( {\kappa \over H} \right)^\prime
       + \left( 2 + {\dot H \over H^2}
       - {F_{,R} \over F} R^\prime \right) \left( {\kappa \over H} \right)
       = - 3 {F_{,R} \over F} R \Bigg[
       \left( {\delta R \over R} \right)^\prime \nonumber \\
   &&  + \left( - {f \over 6 H^2 F}
       + {\mu_m \over 3 H^2 F}
       + {F \over 6 H^2 F_{,R}}
       - {F_{,R} \over F} R^\prime
       + {F_{,RR} \over F_{,R}} R^\prime
       + {R^\prime \over R}
       + {1 \over 3} {k^2 \over a^2 H^2} \right)
       \left( {\delta R \over R} \right) \Bigg]
       + {1 \over H^2 F} \delta \mu_m,
   \label{eq-fR-2}
\eea
\end{widetext}
\bea
   & & \delta_c^\prime
       = {\kappa \over H},
   \label{eq-fR-3} \\
   & & \delta_w^\prime
       = \left( 1 + w \right) \left( {\kappa \over H}
       - {k \over aH} v_w \right),
   \label{eq-fR-4} \\
   & & v_w^\prime
       + \left( 1 - 3 w \right) v_w
       = {w \over 1 + w} {k \over aH} \delta_w,
   \label{eq-fR-5}
\eea where $\delta \mu_m$ and $\delta p_m$ are collective perturbed
density, pressure, respectively. In our full numerical treatment the
collective fluids includes the CDM, baryon, radiation (photons and
neutrinos), etc; the radiation parts include photons handled by
Boltzmann equations or tight coupling approximation. In Eqs.\
(\ref{eq-fR-4}) and (\ref{eq-fR-5}), as an example, we present a
fluid with a constant equation of state parameter $w \equiv
p_w/\mu_w$; $\delta_w=\delta\mu_w /\mu_w$ is the density contrast
and $v_w$ is the perturbed velocity of the dominant fluid. Here a
prime indicates a derivative with respect to $\ln a$ ($F' \equiv
dF/d\ln a$), $k$ is the comoving wave number, and $\kappa$ is the
perturbed expansion of normal frame vector. As we consider the CCG
we need to include the CDM component even in the case it is
subdominant. In case it is subdominant, we have at least a sixth
order differential equation as presented in Eqs.\
(\ref{eq-fR-1})-(\ref{eq-fR-5}).

\section{Initial Conditions for Scaling Evolution}
\label{sec:initial}

In the early era, let us consider a functional form
\begin{equation}
   f(R) = \alpha R^{1+\epsilon},
\label{eq:scl-fR}
\end{equation}
where $\epsilon$ is small positive constant; for generality we
introduced a coefficient $\alpha$ which is unity in our unit. The
gravity of pure power-law form allows scaling evolution in which the
density of $X$-component follows that of the dominant fluid
\cite{Amendola-etal-2007}, and the corresponding potential in the
Einstein frame is a pure exponential potential
\cite{Amendola-etal-2007-PRL}. Here we put an ansatz that $F\mu_X$
evolves as the dominant ideal fluid with constant equation of state
($w=p_w/\mu_w=\delta p_w/\delta\mu_w$) as
\begin{equation}
   F\mu_X = \frac{1}{2}(FR-f)-3H\dot{F}\equiv A\mu_w,
\label{eq:scl-FmuX}
\end{equation}
where $A$ is the constant density fraction of $X$-component relative to the
dominant fluid. By combining Eqs.\ (\ref{eq:ddF}), (\ref{eq:H2}), and
(\ref{eq:scl-FmuX}), we can derive
\begin{eqnarray}
   && \mskip-48mu 3FR-f=2\mu_w [(1-3w)+(2-3w)A], \label{eq:scl-RA} \\
   && \mskip-48mu\frac{F'}{F}
     =\frac{(1-3w)[(1+A)f-FR]+A(1+3w)FR}{(1+A)(f-3FR)}.  \label{eq:scl-dFF}
\end{eqnarray}
Equations (\ref{eq:scl-RA}) and (\ref{eq:scl-dFF}) provide the
scaling initial conditions for $F$ and $F'$, respectively. Noting
that $3FR-f\propto a^{-3(1+w)}$ in Eq.\ (\ref{eq:scl-RA}) and
specifying the form of $f(R)$ in Eq.\ (\ref{eq:scl-fR}), we obtain an
exact expression for $R$
\begin{equation}
   R= H_0^2 \left[\frac{6\Omega_{w0}[(1-3w)+(2-3w)A]}
      {\alpha (2+3\epsilon)(a/a_0)^{3(1+w)}}\right]^{\textstyle \frac{1}{1+\epsilon}},
\label{eq:scl-Rhat}
\end{equation}
and its time derivative
\begin{equation}
   R' = - 3 \frac{1+w}{1+\epsilon} R, \label{eq:scl-dR}
\end{equation}
where $\Omega_{w0}=\mu_{w0}/(3H_0^2)$ is the density parameter of $w$-fluid
at the present epoch (indicated by the subscript $0$).
Inserting Eq.\ (\ref{eq:scl-dR}) into Eq.\ (\ref{eq:scl-dFF}), we determine
\begin{equation}
   A=\frac{\epsilon(7+10\epsilon)+3\epsilon(1+2\epsilon)w}
          {2-3\epsilon-8\epsilon^2-3\epsilon(1+2\epsilon)w}.
\label{eq:scl-A}
\end{equation}
Equations (\ref{eq:scl-Rhat})--(\ref{eq:scl-A}) can be used as
initial conditions for the scaling density evolution of
$X$-component in the $f(R)=R^{1+\epsilon}$ gravity. The scaling
behavior is possible for general constant $w$ value of the dominant
fluid. As we set the scaling initial condition, the density of
$X$-component follows (scales) the dominant fluid component even for
changing $w$ value of the dominant component; for example, from
radiation dominated era to the matter dominated era.

Now, for the perturbed initial conditions, applying the same method used
in Ref.\ \cite{HN-scaling-2001} that presents initial conditions of perturbed
variables during the scaling regime of the cosmology based on a minimally
coupled scalar field, we find initial conditions of perturbed variables in our
$f(R)$ gravity. In the large-scale limit ($\frac{k}{aH} \to 0$),
with the help of Eqs.\ (\ref{eq:H2}), (\ref{eq:scl-fR})--(\ref{eq:scl-A}),
all the background-related coefficients
in Eqs.\ (\ref{eq-fR-1})--(\ref{eq-fR-5}) are expressed in terms of $\epsilon$
and $w$ alone during the scaling regime.  We have solutions for
Eqs.\ (\ref{eq-fR-1})--(\ref{eq-fR-5}) in the form
\bea
   \delta_w\propto \frac{\delta R}{R} \propto e^{n\ln a},
\eea where
\begin{equation}
\begin{split}
   n &=\frac{1-2\epsilon+3w}{1+\epsilon}, \quad
      -\frac{3}{2}\left(\frac{1+w}{1+\epsilon}\right), \\
     & \mskip+24mu \frac{1}{4\epsilon(1+\epsilon)}
        \Bigg\{-3\epsilon+3\epsilon(1+2\epsilon)w
       \pm \bigg[ -16\epsilon  \\
     & \mskip+24mu -31\epsilon^2+160\epsilon^3+256\epsilon^4
        +3\epsilon (16-22\epsilon     \\
     & \mskip+24mu -28\epsilon^2+64\epsilon^3)w
        -9\epsilon^2 (7+12\epsilon-4\epsilon^2)w^2
         \bigg]^{\frac{1}{2}} \Bigg\}.
\end{split}
\label{eq:n}
\end{equation}
The solutions have been obtained by MAPLE software of version 11 \cite{MAPLE}.
By choosing the first one as the growing mode solution, we get initial
conditions of the perturbed variables,
\begin{equation}
\begin{split}
   & \delta_w=C e^{\frac{1-2\epsilon+3w}{1+\epsilon}\ln a}, \quad
     \left(\frac{\kappa}{H}\right)
     =\frac{1-2\epsilon+3w}{(1+\epsilon)(1+w)} \delta_w, \\
   & \frac{\delta R}{R}=\frac{1-4\epsilon-3(1+2\epsilon)w}
       {1+7\epsilon-12\epsilon^2-3(1-\epsilon)(1-2\epsilon)w}\delta_w, \\
   & \mskip-12mu \left(\frac{\delta R}{R}\right)'
     =\frac{1-2\epsilon+3w}{1+\epsilon} \left(\frac{\delta R}{R}\right),
\end{split}
\label{eq:scaling_init_pert}
\end{equation}
where $C$ is the initial amplitude.

Solutions similar to Eq.\ (\ref{eq:n}) were presented in Ref.\
\cite{Carloni-etal-2008}. Compared with the solutions in Ref.\
\cite{Carloni-etal-2008} which were based on the comoving gauge of
the dominant fluid (the $w$-fluid), our solutions are based on the
CCG gauge condition which is the CDM comoving gauge. A comparison of
the two results shows that the growing solution coincides and the
other three solutions differ for $w \neq 0$. We are interested only
in the growing solution, and the solutions in Eq.\
(\ref{eq:scaling_init_pert}) coincide with the one in Ref.\
\cite{Carloni-etal-2008}.

\section{A dark energy model and constraints}
\label{sec:appl}

In this section, we present numerical evolution of background and
perturbed quantities in a dark energy model based on the $f(R)$
gravity with early scaling era. We consider a gravity with a form
\begin{equation}
    f(R)=R^{1+\epsilon}+qR^{-n},
\label{eq:fR_epsn}
\end{equation}
where $\epsilon>0$ and $-1<n\le 0$. Our model allows exact scaling
during the radiation and matter dominated eras (provided by the
first term) and drives the late-time acceleration in the recent
epoch (provided by the second term). Our model with the scaling
initial conditions does not cause the numerical problem discussed in
Sec.\ \ref{sec:intro}. In the early radiation dominated epoch (e.g.,
starting from $a_i/a_0=10^{-11}$ in this paper), for small
$\epsilon=10^{-8} \sim 10^{-6}$ the quantity $F \simeq
(1+\epsilon)R^{\epsilon}$ evolves with values slightly larger than
unity so that Eq.\ (\ref{eq:ddF}) or (\ref{eq:ddR}) can be
numerically manageable within the precision of usual computing
environment. The evolution of both background and perturbed
variables becomes numerically unstable as $\epsilon$ gets smaller
(e.g., $\epsilon \lesssim 10^{-9}$ for $a_i/a_0 = 10^{-11}$). This
is because at the early epoch the quantity $F$ is so close to the
unity that the time-variation of $F$ is not numerically manageable
even in double precision accuracy.

Although the detailed study is not given here, the value of
$\epsilon$ can be more tightly constrained by the solar system test.
For example, according to the criterion given by Ref.\
\cite{Lin-etal-2010}, one expect $\epsilon \lesssim 10^{-17}$ for
$R/H_0^2=10^5$. As we explained earlier, such a small value of
$\epsilon$ is numerically problematic in the early era due to
limited numerical precision currently available. Thus although the
values of $\epsilon$ we consider are likely to be too large
considering the solar-system constraint, as shown below, the case of
$\epsilon=10^{-7}$ and $n=0$ gives power spectra and perturbation
growth that are observationally indistinguishable from the
predictions of the $\Lambda\textrm{CDM}$ model in Einstein gravity
($\epsilon=0$, $n=0$).

There is one numerical task needed during the background evolution
of $f(R)$ gravity. Evolving Eq.\ (\ref{eq:ddF}) or (\ref{eq:ddR})
demands a fine-tuning process to satisfy the condition that the
normalized Hubble parameter at the present epoch should be equal to
unity, $\hat{H}_0=1$, where $\hat{H}\equiv H/H_0$. When the
radiation and the matter energy densities at the present epoch are
fixed, one of the coefficients appearing in the $f(R)$ functional
form should be adjusted to match this condition for given initial
conditions of $R$ and $R'$, or vice versa. In our case we adjust $q$
appearing in Eq.\ (\ref{eq:fR_epsn}) to satisfy $\hat{H}_0=1$ for a
given set of scaling initial conditions. The same numerical
situation also appears in the Einstein gravity with dark energy
model based on a minimally coupled scalar field.

In a modified gravity with the form $f(R)=R+qR^{-n}$, the method of
imposing initial conditions for the background evolution used in the
literature is unclear to us.
Furthermore, the coefficient $q$ is not single-valued depending on
the choice of $n$ and initial conditions of $R$ and $R'$. In an
extreme case when $n=-0.97$, $R_i=10^{-5}H_0^2$, and $R_i'=0$ at
$a_i/a_0=10^{-11}$ with other parameters fixed with the Wilkinson
Microwave Anisotropy Probe (WMAP) 7-year best-fit parameters (see
below), there are nine multiple values of $q$ satisfying
$\hat{H}_0=1$ in the range $-0.7222 < q/H_0^{2n+2} < 0$.
Under the same (non-scaling) condition, multiple $q$ values are also
obtained in our $f(R)$ gravity. On the other hand, our initial
conditions for the scaling evolution presented in Sec.\
\ref{sec:initial} are exact and general so that they can be applied
to any dominant fluid with a constant equation of state $w$. With
the scaling initial conditions,
$q$ is single-valued over the whole range of $n$ considered.

In our analysis we considered cases where the behavior of our $f(R)$
gravity, Eq.\ (\ref{eq:fR_epsn}) can be approximated by the first
term ($R^{1+\epsilon}$; higher power of $R$) at early times and is
dominated by the second term ($qR^{-n}$, $-1<n \le 0$; lower power
of $R$) at late time during the acceleration phase. However, it
should be emphasized that the assumption that the higher power of
$R$ should dominate at early epoch and the lower power of $R$ should
dominate at later epoch is not necessarily the case due to the
nonlinearities of the $f(R)$ gravity theory \cite{Clifton-2008}.
Because the scaling background evolution guaranteed by the scaling
initial condition leads to decreasing $R$ value in time for $w>-1$
[this follows from Eq.\ (\ref{eq:scl-Rhat})], this allows the
transition from higher to lower $R$-power regimes in our case.

\begin{figure}
\includegraphics[width=8.7cm]{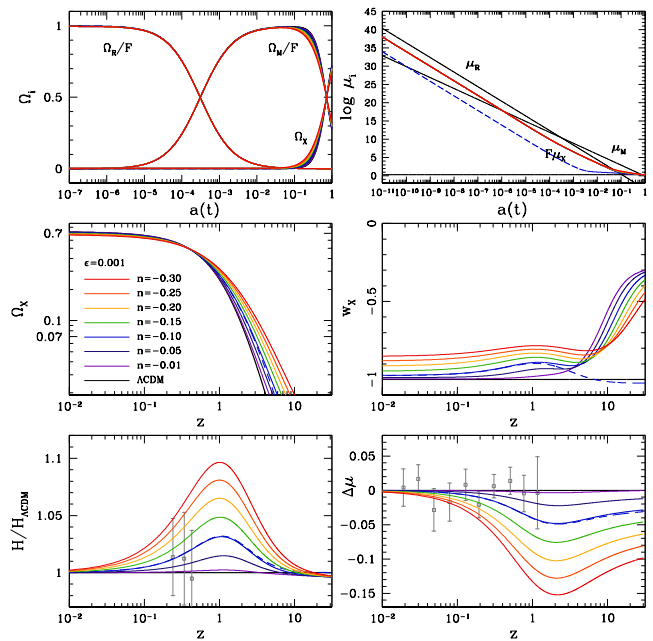}
\caption{Background evolution in the $f(R)=R^{1+\epsilon}+qR^{-n}$ gravity
         with $\epsilon=0.001$ and varying $n=-0.01$, $-0.05$, $-0.10$,
         $-0.15$, $-0.20$, $-0.25$, and $-0.30$ (color solid curves).
         Results for $\epsilon=10^{-7}$ and $n=-0.1$ are added with
         blue dashed curves.
         (Top panels) Evolution of $\Omega_i$ and $\mu_i$ as a function of
         scale factor $a(t)$, where $i=R$, $M$, $X$ indicates radiation,
         matter (baryon plus CDM), and gravity sector, respectively.
         (Middle and bottom panels). Evolution of $\Omega_X$, equation of
         state of $X$-component $w_X$,
         Hubble parameter $H/H_{\Lambda\textrm{CDM}}$, and distance
         modulus $\Delta\mu=\bar{\mu}-\bar{\mu}_{\Lambda\textrm{CDM}}$
         relative to the $\Lambda\textrm{CDM}$ model, where $\bar{\mu}$
         represents the distance modulus.
         In all panels $\Lambda\textrm{CDM}$ predictions are shown as
         thick black curves.
         The Hubble parameter data is taken from Ref.\
         \cite{Gaztanaga-etal-2009}, and the binned SNIa data points
         are from the UNION2 sample \cite{Amanullah-etal-2010}.
         }
\label{fig:bg_fR}
\end{figure}

Figure \ref{fig:bg_fR} shows background evolution of our $f(R)$
gravity with $\epsilon=0.001$ and varying $n$ (color solid curves).
The background parameters shown are density parameters $\Omega_{i}$,
energy densities $\mu_i$, equation of state $w_X$, relative Hubble parameter
($H/H_{\Lambda\textrm{CDM}}$) and distance modulus
($\bar{\mu}-\bar{\mu}_{\Lambda\textrm{CDM}}$) with respect to the fiducial
$\Lambda\textrm{CDM}$ model.
As a fiducial model we take a flat $\Lambda$CDM universe with parameters
$\Omega_{M0}=0.274$ ($\Omega_{c0}=0.2284$ and $\Omega_{b0}=0.0456$),
$\Omega_{\Lambda 0}=0.7278$, $h=0.704$, $n_s=0.963$, $\sigma_8=0.809$,
$T_0=2.725~\textrm{K}$, $Y_\textrm{He}=0.24$, $N_\nu=3.04$ with
reionization optical depth $\tau=0.087$ based on the WMAP 7-year observations
\cite{WMAP7}.
In all $f(R)$ gravity models, we set $\Omega_{X0}\equiv \Omega_{\Lambda 0}$.
With the fiducial model parameters, the $\epsilon$ has an upper bound
$\epsilon_\textrm{BBN}=0.011314$ so that the initial contribution from
the dark energy is lower than the maximum amount allowed by the
big bang nucleosynthesis (BBN) calculation, $\Omega_{Xi} < 0.045$
\cite{Bean-etal-2001}.
We also added blue dashed curves in Fig.\ \ref{fig:bg_fR} to represent the
background evolution for $\epsilon=10^{-7}$ and $n=-0.1$, demonstrating that
the energy density of $X$-component (here $F\mu_X$) becomes less
dominant than the dominant fluids ($\mu_R$ and $\mu_M$) as $\epsilon$
gets smaller.
The background observables (Hubble parameter and distance modulus)
for $\epsilon < 10^{-3}$ are largely insensitive to $\epsilon$ and
are very similar to those for $\epsilon=0.001$.

\begin{figure}
\includegraphics[width=8.7cm]{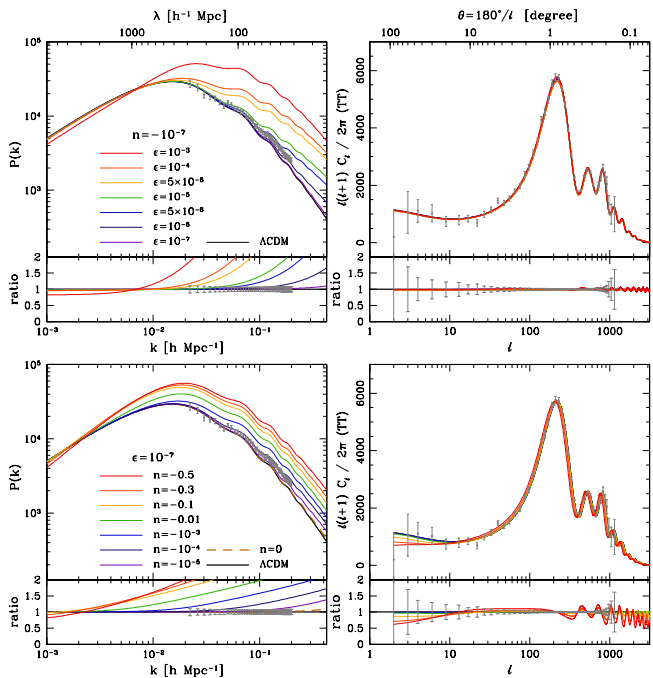}
\caption{Top panels: baryonic matter (left) and CMB temperature anisotropy
         (right) power spectra in the $f(R)=R^{1+\epsilon}+qR^{-n}$ gravity
         models with $n=-10^{-7}$ and varying $\epsilon=10^{-7}$, $10^{-6}$,
         $5\times 10^{-6}$, $10^{-5}$, $5\times 10^{-5}$, $10^{-4}$, and
         $10^{-3}$.
         Bottom panels: The same as in the top-panels but with
         $\epsilon=10^{-7}$ and varying $n=-10^{-5}$, $-10^{-4}$,
         $-10^{-3}$, $-0.01$, $-0.1$, $-0.3$, and $-0.5$.
         The results for $\epsilon=10^{-7}$ and $n=0$
         are shown as brown dashed curves, which are very similar to
         the $\Lambda\textrm{CDM}$ predictions (black curves).
         In all models including $\Lambda\textrm{CDM}$ model,
         the same initial amplitude has been assumed.
         The ratios of $f(R)$ gravity power spectrum to $\Lambda\textrm{CDM}$
         prediction are also shown in the bottom region of each panel.
         For matter and CMB power spectra, recent measurement from
         the Sloan Digital Sky Survey (SDSS) DR7 luminous red galaxies (LRG)
         \cite{SDSS-DR7-LRG}
         and the WMAP 7-year data (including the cosmic variance)
         \cite{Larson-etal-2010} have
         been added (grey dots with error bars) together with fractional
         errors of observed spectra.
         }
\label{fig:fR_pow}
\end{figure}
\begin{figure}
\includegraphics[width=8.7cm]{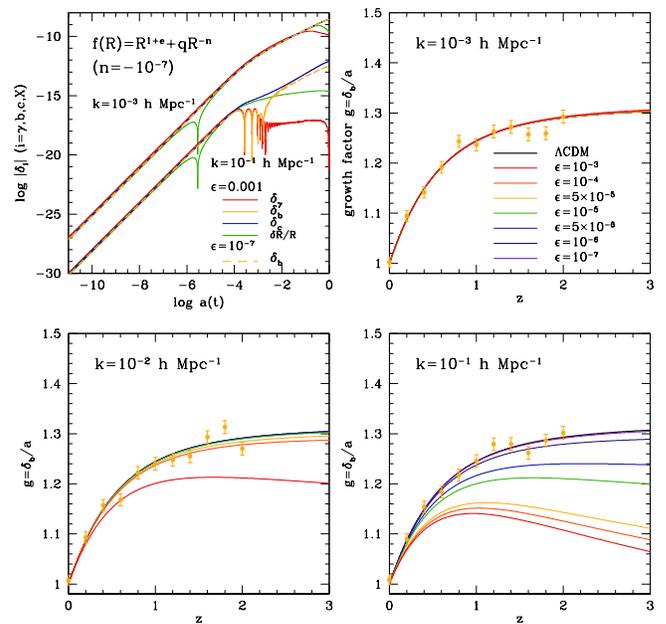}
\caption{Top-left: evolution of density perturbations $\delta_i$
         ($i=\gamma,b,c,X$) at comoving scales $k=10^{-3}$ and
         $10^{-1}~h\textrm{Mpc}^{-1}$ in $f(R)=R^{1+\epsilon}+qR^{-n}$
         gravity with $n=-10^{-7}$ for $\epsilon=0.001$ (solid) and
         $\epsilon=10^{-7}$ (dashed curves).
         Perturbation growth of photon ($\gamma$), baryon ($b$), CDM ($c$),
         and $\delta R/R$ are represented as red, yellow, blue, green curves,
         respectively.
         Other panels: evolution of growth factor $g\equiv\delta_b/a$
         (normalized to unity at present)
         at comoving scales $k=10^{-3}$, $10^{-2}$, and
         $10^{-1}~h\textrm{Mpc}^{-1}$ for models with parameters used
         in the top-panels of Fig.\ \ref{fig:fR_pow}
         ($n=-10^{-7}$ and varying $\epsilon$) with the same colored code.
         Growth factors of the $\Lambda\textrm{CDM}$ model are shown as
         black curves. Note that since we consider interactions between
         radiation and baryon components, the $\Lambda\textrm{CDM}$ growth
         factor is mildly scale-dependent.
         Yellow dots with error bars indicate the $\Lambda\textrm{CDM}$
         growth factor expected from the future $X$-ray and weak-lensing
         observations \cite{Vikhlinin-etal-2009}.
         }
\label{fig:fR_ggg_n7}
\end{figure}

\begin{figure}
\includegraphics[width=8.7cm]{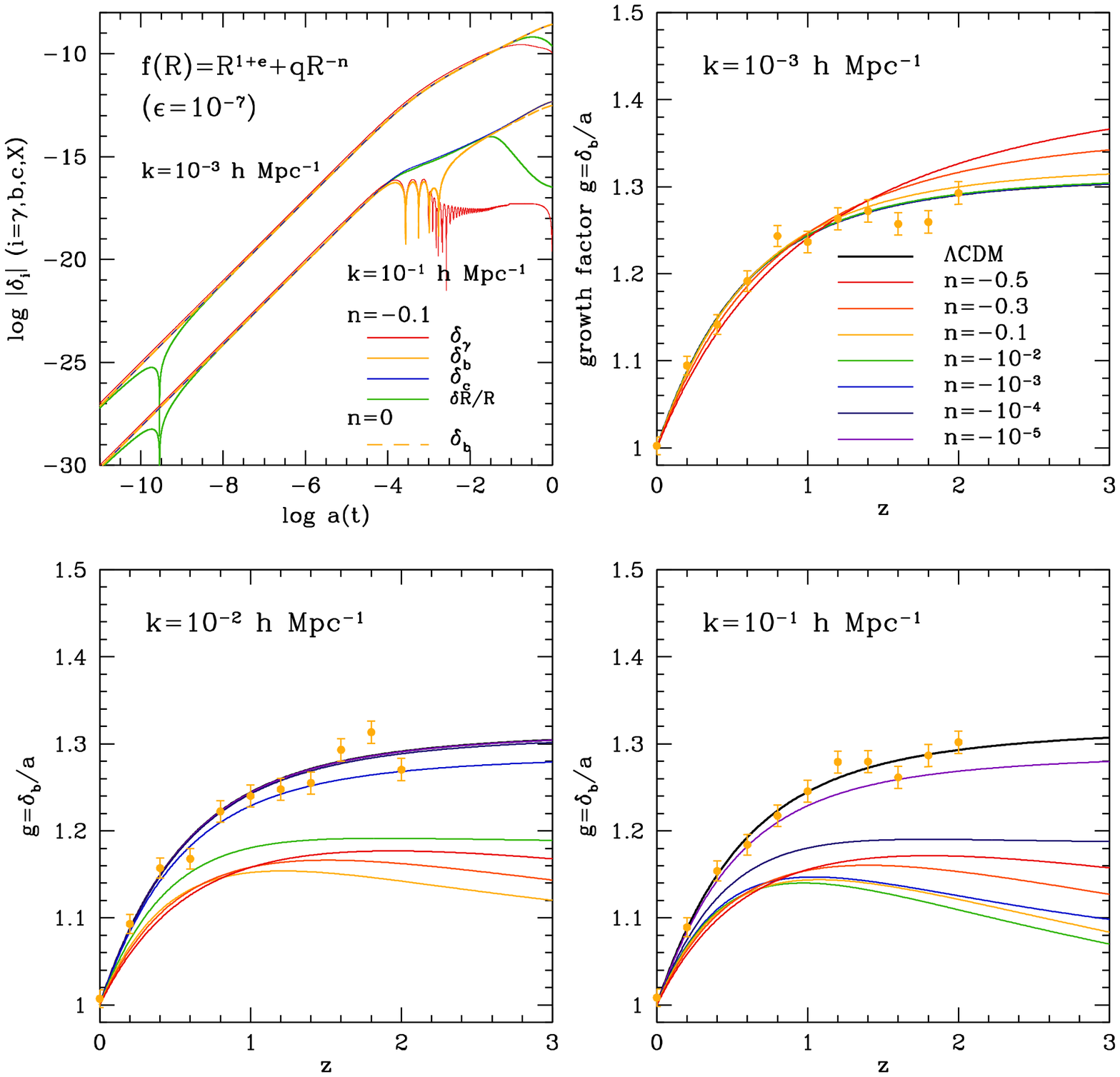}
\caption{Same as Fig.\ \ref{fig:fR_ggg_n7} but for $f(R)$ gravity models
         ($\epsilon=10^{-7}$ and varying $n$)
         used in the bottom-panels of Fig.\ \ref{fig:fR_pow} with the same
         colored code.
         }
\label{fig:fR_ggg_e7}
\end{figure}

In order to calculate the matter and CMB power spectra,
we solve a system composed of matter (baryon [$b$] and CDM [$c$]),
radiation (photon [$\gamma$] and neutrinos [$\nu$]), and the $f(R)$
gravity sector as dark energy.
The perturbation in the radiation component has been handled by using the
Boltzmann equation or tight coupling approximation (see Ref.\
\cite{HN-CMB-2002} for our set of equations and the numerical methods).
As the initial conditions of background and perturbation variables we use
Eqs.\ (\ref{eq:scl-Rhat})--(\ref{eq:scl-A}) and (\ref{eq:scaling_init_pert})
with $w=\frac{1}{3}$ and $\Omega_{w0}=\Omega_{R0}$.
All the perturbed variables we use are spatially gauge invariant
\cite{Bardeen-1988}.
We solved the system by adopting the CCG as the temporal gauge condition.
The CCG is the same as the synchronous gauge without the gauge mode.
For the matter power spectrum we present the power spectrum of baryonic
density perturbation based on the CCG which is a gauge-invariant concept.
The CMB anisotropy is naturally gauge invariant.
Figure \ref{fig:fR_pow} shows the matter and CMB temperature anisotropy
power spectra for $n=-10^{-7}$ with varying $\epsilon$ (top) and
for $\epsilon=10^{-7}$ with varying $n$ (bottom panels).
We omit CMB polarization power spectra.
Note that the baryonic matter power spectra is very sensitive to both
$\epsilon$ and $n$ while the CMB power spectra (including polarization)
are mildly sensitive to the variation of $n$ and are insensitive to $\epsilon$.

Our results for $\epsilon=10^{-7}$ and $n=0$ (shown as brown dashed
curves in the bottom panels of Fig.\ \ref{fig:fR_pow}) are very
similar to the $\Lambda\textrm{CDM}$ predictions (black curves); the
deviations are less than $2$\% level at scales $k <
0.2~h\textrm{Mpc}^{-1}$ for matter power spectrum and less than
$0.5$\% level for CMB power spectrum. Our calculations of power
spectra for $\epsilon=10^{-7}$ and nonzero $n$ are consistent with
those presented in Ref.\ \cite{Li-Barrow-2007} that considered
$f(R)=R+qR^{-n}$ gravity. We have also checked that the power
spectra for $\epsilon=10^{-7}$ are indistinguishable within the
$1$\% precision from those obtained by evolving the perturbed
variables under the model which uses the $\Lambda\textrm{CDM}$ model
in Einstein gravity at the early epoch ($a/a_0 < 0.001$) and
suddenly switches to the $f(R)=R+qR^{-n}$ gravity thereafter.

Figure \ref{fig:fR_ggg_n7} displays the perturbation growth of
individual components ($\delta_\gamma$, $\delta_b$, $\delta_c$,
$\delta R/R$) and baryon perturbation growth factor $g\equiv
\delta_b /a$ at recent epoch in our $f(R)$ gravity with $n=-10^{-7}$
and varying $\epsilon$. We see that each perturbed variable follows
with each other at the early epoch, which demonstrates the scaling
evolution. In the growth factor panels we have added
$\Lambda\textrm{CDM}$ mock growth factor data points that are
expected from the future X-ray and weak-lensing observations
\cite{Vikhlinin-etal-2009}. The mock data points, located in the
redshift range $z=0$--$2$ with equal interval of $\Delta z=0.2$,
have been generated by adding $1$\% random noise to the
$\Lambda\textrm{CDM}$ growth factor at comoving scales $k=0.001$,
$0.01$, and $0.1~h\textrm{Mpc}^{-1}$. The perturbation growth factor
at the small scale ($k=0.1~h\textrm{Mpc}^{-1}$) is very sensitive to
the variation of $\epsilon$. It is also sensitive to the variation of $n$,
which is shown in Fig.\ \ref{fig:fR_ggg_e7}.

\begin{figure}
\includegraphics[width=8.7cm]{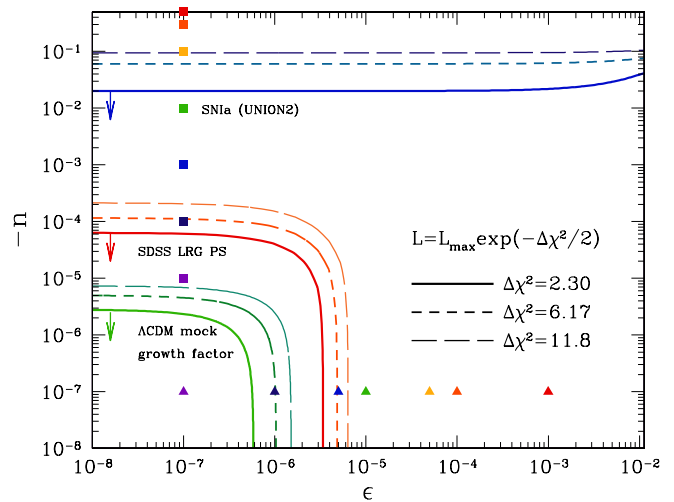}
\caption{Likelihood of $f(R)$ gravity parameters constrained by
         the UNION2 SNIa sample (blue contours),
         the SDSS DR7 LRG power spectrum (red contours), and
         the $\Lambda\textrm{CDM}$ mock growth factor data
         at $k=0.1~h\textrm{Mpc}^{-1}$ (green contours).
         The levels of contours have been determined by setting
         $\Delta\chi^2=2.30$ (solid), $6.17$ (short-dashed), $11.8$
         (long-dashed curves) relative to the $\chi^2$-minimum,
         mimicking $1\sigma$, $2\sigma$, $3\sigma$ confidence levels,
         respectively.
         Arrows indicate directions to which the likelihood increases.
         Triangles and squares indicate the models shown in top and bottom
         panels of Fig.\ \ref{fig:fR_pow} (Fig.\ \ref{fig:fR_ggg_n7}
         and Fig.\ \ref{fig:fR_ggg_e7} for the perturbation growth rate),
         respectively, with the same colored code.
         }
\label{fig:fR_like}
\end{figure}

From Figs.\ \ref{fig:fR_pow}-\ref{fig:fR_ggg_e7} we notice that the
CMB power spectrum is less sensitive to the model parameters than
the baryon density power spectrum and the baryon growth rate. Thus,
in the following we use the baryon density power spectrum and the
baryon growth rate together with the SNIa data to constrain the
model parameters. Figure \ref{fig:fR_like} shows likelihood
distributions of $f(R)$ gravity parameters constrained by the recent
UNION2 SNIa data set (without systematic errors)
\cite{Amanullah-etal-2010}, the SDSS DR7 LRG power spectrum (PS)
\cite{SDSS-DR7-LRG}, and the $\Lambda\textrm{CDM}$ mock growth
factor data at the comoving scale $k=0.1~h\textrm{Mpc}^{-1}$ (in the
bottom-right panels of Figs.\ \ref{fig:fR_ggg_n7} and \ref{fig:fR_ggg_e7}).
Our likelihood estimation is tentative since we have explored only the
$\epsilon$--$n$ space while other cosmological parameters are fixed
with the WMAP 7-year best-fit parameters. The likelihood
distribution has been calculated from $L=L_{\textrm{max}}
\exp(-\Delta\chi^2/2)$, where $L_\textrm{max}$ has been taken as the
maximum value within the parameter space we have explored, $10^{-8}
\le \epsilon \le \epsilon_\textrm{BBN}$ and $-0.5 \le n \le
-10^{-8}$. The $1\sigma$, $2\sigma$, $3\sigma$ confidence levels
have been roughly determined by setting $\Delta\chi^2=2.30$, $6.17$,
$11.8$ from the $\chi^2$-minimum, which is valid only for the
Gaussian distribution. In the $\chi^2$-estimation with the SDSS LRG
power spectrum, we consider the convolution effect caused by
band-power window functions and exclude data points where the
non-linear clustering dominates (see Appendix C of Ref.\
\cite{Park-etal-2010} for detailed description). As shown in Fig.\
\ref{fig:fR_like}, $f(R)$ gravity parameters, $\epsilon$ and $n$,
 are very sensitive to the growth factor at small
scales, and are already tightly constrained by the current
measurement of galaxy power spectrum. For $\Lambda\textrm{CDM}$ mock
growth factor data, the likelihood distribution suggests that only
the parameter space extremely close to the $\Lambda\textrm{CDM}$
model is allowed.

\section{Discussion}

In this paper we have studied a $f(R)$-gravity based dark energy model
with early scaling era. We have presented initial conditions of background
and perturbed variables during the early scaling evolution regime in
the modified gravity with a pure power-law form $f(R)=R^{1+\epsilon}$
in the early era. With these initial conditions, the modified gravity
with a form $f(R)=R^{1+\epsilon}+f_\textrm{DE}(R)$ where the second term
drives the late-time acceleration becomes free from the numerical
difficulty that is usually confronted during the background
evolution in the early radiation dominated era for
$f(R)=R+f_\textrm{DE}(R)$ gravity. Our initial conditions are
general so that the scaling density evolution of the $X$-component
is assured for any dominant fluid with a constant equation of state
parameter $w$.

As a possible dark energy model we have considered the gravity with
a form $f(R)=R^{1+\epsilon}+qR^{-n}$ and compared the evolution of
the background and perturbation variables in this gravity with the
recent observational data and the $\Lambda\textrm{CDM}$ mock data.
The present observational data already severely constrain our model
parameters $n$ and $\epsilon$ so that only parameters extremely
close to the $\Lambda\textrm{CDM}$ model is allowed. We found that
the power spectrum of baryon component and the perturbation growth
factor at small scales are more sensitive to the $f(R)$ gravity
parameters than the SNIa distance modulus and the CMB anisotropy
power spectra (see Figs.\ \ref{fig:fR_pow}--\ref{fig:fR_like}).
Therefore, precise measurement of the perturbation growth is
essential to tightly constrain our $f(R)$ gravity.

%
%
\acknowledgments
We thank Dr.\ Yong-Seon Song for useful discussions. We also wish to
thank the anonymous referee for the constructive and helpful
comments on our manuscript. H.N.\ was supported by Mid-career
Research Program through National Research Foundation funded by the
MEST (No.\ 2010-0000302). J.H.\ was supported by the Korea Research
Foundation Grant funded by the Korean Government
(KRF-2008-341-C00022).

\def\and{{and }}


\end{document}